\documentclass[superscriptaddress,aps,pra,10pt,twocolumn,showpacs,notitlepage,nofootinbib,longbibliography]{revtex4-1}
\usepackage{etex}
\usepackage{amsmath,amssymb,amsthm}
\usepackage[colorlinks=true,citecolor=blue,urlcolor=blue]{hyperref}
\usepackage[pdftex]{graphicx}
\usepackage{times,txfonts}
\usepackage{braket}
\usepackage{color}
\usepackage{natbib}
\usepackage{amsmath,blkarray}
\usepackage{mathtools}
\usepackage{latexsym}
\usepackage{tabularx, booktabs}
\usepackage{graphics,epstopdf}
\usepackage{graphicx}
\usepackage{float}
\usepackage{graphicx}
\usepackage{graphicx}
\usepackage{amsfonts}
\usepackage{subcaption}
\usepackage{multirow}
\usepackage{amsmath}
\usepackage{amsthm}
\usepackage[toc]{appendix}

\newcommand{\be}{\begin{equation}}
\newcommand{\ee}{\end{equation}}
\newcommand{\ba}{\begin{eqnarray}}
\newcommand{\ea}{\end{eqnarray}}

\allowdisplaybreaks
\usepackage[dvipsnames]{xcolor}

\usepackage{titlesec}
\usepackage{comment}
\begin{document}

\title{Activation of entanglement in generalized entanglement swapping}
\author{Pratapaditya Bej}
\email{pratap6906@gmail.com }
\affiliation{Department of Physics and Center for Astroparticle Physics and Space Science, Bose Institute, EN Block, Sector V, Saltlake, Kolkata - 700091, India}

\affiliation{ExamRoom.AI, No: 291-92/1A, Venkatadri Building, Ground floor, Konappa Agrahara, Electronic City Phase 1, Bengaluru - 560100, India }

\author{Abhishek Banerjee}
\email{banerjee.abhishek.official@gmail.com}
\affiliation{Department of Physics and Center for Astroparticle Physics and Space Science, Bose Institute, EN Block, Sector V, Saltlake, Kolkata - 700091, India}

\begin{abstract}
We study entanglement activation in a generalized entanglement swapping process involving two Bell pairs and generalized measurements. The conventional understanding posits entangled measurements as both necessary and sufficient for establishing entanglement between distant parties. In this study, we reassess the role of measurement operators in entanglement generation within a generalized entanglement swapping process. We focus on maximally entangled two-qubit initial states and generalized
measurements, investigating the necessity and sufficiency conditions for entangled measurement operators. By
utilizing two Bell pairs, (1, 2) shared between Alice and Bob, and (3, 4) shared between Bob and Charlie,
we demonstrate that while entangled measurements are sufficient, they are not indispensable for establishing
entanglement between spatially separated observers. Through a sequential approach, if Bob performs an initial measurement which is not able to establish entanglement then followed by another measurement after post-processing the first measurement it is possible to establish entanglement. We identify specific criteria for different measurement operators that enable the potential for performing a second measurement to establish entanglement. Our findings highlight the feasibility of generating entanglement between distant parties through a combination of measurements, shedding  light on entanglement distribution in
quantum networks. Additionally, we showcase through illustrative examples how successive measurements enhance entanglement compared to single measurements, underscoring the practical benefits of our approach in enhancing entanglement. Moreover, our protocol extends beyond bipartite qubit states to higher-dimensional maximally entangled states, emphasizing its versatility and applicability.

\end{abstract}
\maketitle
\section{INTRODUCTION}
Shared entanglement is a valuable resource in quantum information theory, as it facilitates various quantum information theoretic tasks such as Teleportation \cite{tele}, Super dense coding \cite{dense}, among others. Distribution of entanglement in itself is a highly non-trivial task. 

Entanglement swapping \cite{es1,sbose} is one of the most well-known protocols used to distribute entanglement over distant nodes which haven't interacted in the past. In its canonical form, entanglement swapping comprises the following setup and steps. Three spatially separated parties namely Alice (A), Bob (B), and Charlie (C) share entangled states among themselves such that Alice and Bob share an entangled pair of qubits (1, 2). Bob and Charlie share another entangled pair of qubits (3, 4). Qubits (2, 3) are in Bob's possession. The goal is to establish entanglement between Alice and Charlie (1, 4). This is achieved by performing a measurement in Bell basis on Bob's pair of qubits (2, 3), which leads to Alice and Charlie's qubits (1, 4) being entangled. 

In the original setting, the initially shared entangled states were chosen to be maximally entangled states and the measurement at Bob's location was Bell basis measurement. This can be altered to work in a more general scenario with non-maximally entangled states as well as more general measurement settings \cite{pbej1,pbej2}.

 In operational terms, an operator is referred to as an inseparable operator if its Choi state is entangled while separable operators are operators which take separable states to separable states and its Choi state is a separable density matrix \cite{rain1,rain2,rain3}. Quantum measurements are a special class of operators. The most general form of measurement is called a positive operator-valued measure (POVM) \cite{um1,unsharp_measurement,povm}. A POVM,  $\mathcal{M}$ is defined as a set of positive operators $\Pi_n$ so that $\sum_n{\Pi_n} = \mathbb{I}$ where $\mathbb{I}$ is the identity operator on the corresponding Hilbert space. Measurements are characterized by various definitions. In papers \cite{yokoyama,hamamura,sperling,duan,vertesi,rabelo,adam,wu,virmani}, it has been defined that a POVM element $\Pi_{n}$ is considered entangled if the partial transpose of the corresponding operator, defined as $\frac{\Pi_n}{\mbox{Tr}(\Pi_n)}$, fails to be positive. The POVM element is said to be unentangled if the normalized form $\frac{\Pi_n}{\mbox{Tr}(\Pi_n)}$ is a separable quantum mixed state.  A POVM measurement is called an entangled measurement if at least one of the POVM elements is entangled. Similarly, A POVM measurement is called unentangled measurement if each of the POVM elements is unentangled.  Here, the entanglement of the measurement is characterized as an \textit{entanglement of the detection devices} or POVM entanglement. The entire measurement set is classified into different categories based on this definition \cite{vertesi,adam}. In \cite{vertesi,adam}, the authors have classified the set of measurements into different categories such as entangled measurement, unentangled measurement, local operations and classical communication (LOCC) measurement, and classical measurement where classical and LOCC measurements do not have any capacity to produce the entangled state from a separable state. Note that these measurements have the inclusion relation: General measurement  $\supset$ unentangled measurement $\supset$ LOCC measurement $\supset$ classical measurement, where general measurement corresponds to the set of all quantum measurements, including the entangled measurements \cite{vertesi,adam}.  LOCC and classical measurements belong to the class of separable operations in an operational sense. It is important to note that, according to this definition, not every unentangled measurement is considered a separable operation, but the converse is true.





 The standard entanglement swapping procedure has been generalized to work with non-maximally entangled states, and projective measurements have found important applications in quantum networks and nonlocality-related problems \cite{perseguers1,perseguers2,wojcik,klobus,gour,verma}. Recent studies have proposed elegant protocols leveraging POVM measurements for entanglement distribution in quantum networks \cite{halder1,halder2}. The common understanding suggests that entangled measurements are required for establishing entanglement between distant parties \cite{yokoyama,perseguers2}.

In this work, we re-evaluate the necessity and sufficiency conditions for measurement operators to be entangled when initial states are maximally entangled and measurements are generalized measurements. We start with two Bell pairs: (1, 2) shared between Alice and Bob, and (3, 4) shared between Bob and Charlie. A POVM measurement is applied to qubits (2, 3), leading to the creation of a shared state in (1, 4) between the spatially separated observers. We demonstrate that while an entangled measurement is sufficient, it is not necessary for the measurement operators to be entangled to establish an entangled state between A and C. By employing a sequential approach where Bob conducts an initial measurement, and if that does not establish the entanglement, followed by another measurement, it is possible to achieve entanglement. This sequential strategy, combined with appropriate post-processing after the first measurement, enables the establishment of entanglement between A and C. We identify the specific criteria for different measurement operators that enable the possibility of performing a second measurement to establish entanglement. In cases where the first measurement fails to establish entanglement, we delineate the protocol accordingly. We demonstrate that after the first measurement, the $14|23$ bipartition entanglement of the total system state should be between 0 and 1, and the rank of the measurement should be greater than one to enable a disturbance of the (1,4) state by the second measurement. Additionally, we show that a zero $14|23$ bipartition entanglement resulting from a measurement that is an inseparable operator is not feasible. Our demonstration showcases the feasibility of generating entanglement between A and C for different measurement operators. Furthermore, an illustrative example underscores how successive measurements yield more substantial entanglement compared to a single measurement. Specifically, when employing one unentangled measurement incapable of establishing entanglement, a second measurement involving purely separable operation becomes effective. Our approach is not limited to bipartite qubit states but can also be generalized to higher-dimensional maximally entangled states.

The remainder of the paper is structured as follows: Section \ref{sec1} outlines the protocol and presents the results. Finally, Section \ref{sec2} provides concluding discussions and highlights avenues for future research.


 \section{GENERALIZED PROTOCOL}\label{sec1}
We consider a generalized entanglement-swapping scenario, where Alice and Bob share a  Bell pair 
$|\phi^+\rangle_{12}=\frac{1}{\sqrt{2}}(|00\rangle+|11\rangle)$ denoted by ($1,2$). Similarly, Bob and Charlie share another Bell pair $|\phi^+\rangle_{34}=\frac{1}{\sqrt{2}}(|00\rangle+|11\rangle)$ denoted by ($3,4$). Here Alice, Bob, and Charlie are physically separated from each other. The total system state is 
\begin{equation}
\label{eq1}
|\Phi\rangle=|\phi^+\rangle_{12}\otimes|\phi^+\rangle_{34},
\end{equation}
a pure state which is separable in $12|34$ bipartition. Now Bob's first measurement $\mathcal{M}_1$ can be described by a set of two-qubit positive-semidefinite elements $\{\Pi_n \bigm|n=1,\ldots,K, \;\Pi_n\geq0,\; \sum_{n}\Pi_n=\mathbb{I}_4\}$ where each POVM element $\Pi_n$ can be decomposed in the form

\begin{equation}
\label{eq2}
  \Pi_n=\sum_{\alpha=1}^4\pi_{n\alpha}|\phi_{n\alpha}\rangle\langle\phi_{n\alpha}|,
\end{equation}
where $|\phi_{n\alpha}\rangle$, $\alpha=1,2,3,4,$ form a complete orthonormal basis on $\mathbb{C}^2\otimes\mathbb{C}^2$, $\pi_{n\alpha}\geq0$ for $\alpha=1,2,3,4$. Here, $K$ represents the number of POVM elements in a POVM measurement.
If Bob performs a joint two-qubit generalized quantum measurement $\mathcal{M}_1$ on ($2,3$) of the $|\Phi\rangle$ state, then the post-measurement state of the  $n$th outcome  is 
\begin{equation}
    \label{eq3}
    |\Phi_n\rangle=\frac{1}{\sqrt{p_n}}\sqrt{\Pi_n}\otimes \mathbb{I}_4|\Phi\rangle,
\end{equation}
with probability $p_n$ where it is understood that $\Pi_n$ is acting on (2,3) pair and identity operator ($\mathbb{I}_4$) acting on (1,4) pair. The probability of obtaining $n$th outcome is
\begin{eqnarray}
    \label{eq4}
    p_n&=&\langle\Phi|\Pi_n\otimes\mathbb{I}_4|\Phi\rangle\nonumber\\
    &=&\frac{1}{4}\mbox{Tr}(\Pi_n).
\end{eqnarray}

Using Eqs.(\ref{eq1}) and (\ref{eq2}) in (\ref{eq3}) \cite{pbej1}, one can get 
\begin{equation}
    \label{eq5}
    |\Phi_n\rangle=\frac{1}{2\sqrt{p_n}}\sum_{\alpha=1}^4\sqrt{\pi_{n\alpha}}|\phi_{n\alpha}\rangle_{23}|\phi_{n\alpha}^*\rangle_{14},
\end{equation}
where $|\phi^*\rangle$ denotes the complex conjugate of $|\phi\rangle$ in the computational basis. Note that after the measurement, the total system state (\ref{eq5}) remains pure. 

Now after tracing out (2,3) qubits the post-measurement state between Alice and Charlie for $n$th outcome becomes
\begin{eqnarray}
    \label{eq6}
    \rho_{14|n}&=&\frac{1}{4p_n}\sum_{\alpha=1}^4\pi_{n\alpha}|\phi_{n\alpha}^*\rangle\langle\phi_{n\alpha}^*|_{14}\nonumber\\
    &=&\frac{\Pi_n^*}{\mbox{Tr}(\Pi_n)},
\end{eqnarray}
where $\Pi_n^*=\sum_{\alpha=1}^4\pi_{n\alpha}|\phi_{n\alpha}^*\rangle\langle\phi_{n\alpha}^*|$. We also determine the post-measurement state of (1,2) and (3,4) pairs; for details, see Appendix \ref{ap1}. Equation (\ref{eq6}) illustrates how Bob's measurement fully determines the state of the pair (1, 4). This equation also highlights the direct relationship between the post-measurement state of the (1,4) pair and the POVM measurement element.  Conventional thought suggests that entanglement at the (1,4) pair requires an entangled measurement. However, we demonstrate that although an entangled POVM
is sufficient to obtain an entangled state at (1,4), it is not a necessary condition

Our demonstration reveals that through a sequential approach, where Bob performs an initial measurement  $\mathcal{M}_1$ and, if necessary, follows it with another measurement $\mathcal{M}_2$, entanglement can be achieved. This sequential strategy, coupled with appropriate post-processing after the first measurement, facilitates the establishment of entanglement between parties A and C. Notably, the second measurement $\mathcal{M}_2$ does not need to be an inseparable operation; a measurement that is separable operation can also accomplish the task.


Now, we consider that Bob post-processes each outcome of the first measurement $\mathcal{M}_1$ and then performs a second measurement $\mathcal{M}_2$ on the (2,3) qubits of the total system state (\ref{eq5}). Additionally, Bob classically communicates the outcomes to Alice and Charlie during the measurement process.

We consider the second measurement $\mathcal{M}_2$  is also two qubits generalized measurement (POVM) which can be specified by a collection of two qubits semi-definite operators $E_m$ satisfying $\sum_{m=1}E_m=\mathbb{I}_4$ where each POVM element $E_m$ can be decomposed in the previous form  as
\begin{equation}
    \label{eq13}
    E_m=\sum_{\delta=1}^4\mu_{m\delta}|\psi_{m\delta}\rangle\langle\psi_{m\delta}|,
\end{equation}
 where $|\psi_{m\delta}\rangle$, $\delta=1,2,3,4$, form a complete orthonormal basis on $\mathbb{C}^2\otimes\mathbb{C}^2$, $\mu_{m\delta}\geq0$ for $\delta=1,2,3,4$. 
 
The post-measurement state of the four-qubit system, when Bob's outcomes for the successive measurements are the $n$-th element of the measurement $\mathcal{M}_1$ and the $m$th element of $\mathcal{M}_2$, is given by
 \begin{eqnarray}
     \label{eq14}
     |\Phi_{nm}\rangle=\frac{1}{\sqrt{s_{nm}}}\sqrt{E_m}\otimes \mathbb{I}_4|\Phi_n\rangle
 \end{eqnarray}
where it is understood that $E_m$ is acting on (2,3) pair and $\mathbb{I}_4$ is acting on (1,4) pair of $|\Phi_n\rangle$  which is given by Eq.(\ref{eq5}).  The probability of obtaining the $|\Phi_{nm}\rangle$ as the outcome state is $p_ns_{nm}$.  

Here, we denote by $s_{nm}$ the probability of obtaining the $m$-th outcome of the second measurement, $\mathcal{M}_2$ when it acts on the $n$-th outcome of the first measurement $\mathcal{M}_1$. Using Eqs.(\ref{eq4}), (\ref{eq5}), and (\ref{eq13}), we get 
\begin{eqnarray}
    \label{eq15}
 s_{nm}&=&\langle\Phi_n|E_m\otimes\mathbb{I}_4|\Phi_n\rangle\nonumber\\
 &=&\frac{1}{\mbox{Tr}(\Pi_n)}\sum_{\alpha,\delta=1}^4 \pi_{n\alpha}\mu_{m\delta}|\langle\psi_{m\delta}|\phi_{n\alpha}\rangle|^2. 
\end{eqnarray}

Now using Eqs.(\ref{eq5}), and (\ref{eq13}),  the above Eq.(\ref{eq14}) can be written as
\begin{eqnarray}
\label{eq16}
    |\Phi_{nm}\rangle&=&\frac{1}{2\sqrt{p_ns_{nm}}}\sum_{\alpha,\delta=1}^4\sqrt{\pi_{n\alpha}}\sqrt{\mu_{m\delta}}\langle\psi_{m\delta}|\phi_{n\alpha}\rangle|\psi_{m\delta}\rangle_{23}\otimes|\phi_{n\alpha}\rangle_{14},\nonumber\\
\end{eqnarray}
where $p_n$ and $s_{nm}$ are given by Eqs.(\ref{eq4}) and (\ref{eq15}) respectively. 

 After tracing out the (2,3) pair from the above Eq.(\ref{eq16}), we obtain the post-measurement state between A and C for each outcome, given by 
\begin{widetext}
\begin{eqnarray}
    \label{eq17}
    \rho_{14|nm}=\frac{1}{4p_ns_{nm}}\sum_{\alpha,\beta=1}^4\sum_{\delta=1}^4\mu_{m\delta}\sqrt{\pi_{n\alpha}}\sqrt{\pi_{n\beta}}\langle\phi_{n\alpha}|\psi_{m\delta}\rangle\langle\psi_{m\delta}|\phi_{n\beta}\rangle |\phi_{n\alpha}\rangle\langle\phi_{n\beta}|,
\end{eqnarray}
\end{widetext}
where $|\phi_{n\alpha(\beta)}\rangle$, $\alpha(\beta)=1,2,3,4$, are orthonormal basis of the $n$th element $\Pi_n$ of the first measurement $\mathcal{M}_1$.

As we are solely interested in the post-measurement state between A and C for successive measurements, we did not assess the post-measurement state of the (1,2) and (3,4) pairs for the next round of measurement.

\paragraph*{} 
Next, to analyze the bipartition entanglement of post-measurement states for each outcome, we utilize the I-concurrence \cite{rungta} for higher-dimensional pure quantum systems. Throughout this work, when referring to bipartition entanglement, we specifically mean the I-concurrence. The bipartition entanglement $A|B$ is defined using the following expression:
\begin{eqnarray}
    \label{eq18}
    \mathcal{C}_{AvsB}(\rho_{AB})=\sqrt{\frac{d}{d-1}[1-\mbox{Tr}(\rho_{A}^2)]}
\end{eqnarray}
where $d$ is the dimension of $\rho_{A}$, which is reduced state of  $\rho_{AB}$. 

\paragraph*{}
We have already observed that if $\Pi_n$ of $\mathcal{M}_1$ is entangled, then the corresponding post-measurement state between A and C is always entangled. Now, instead of an entangled measurement, if $\Pi_n$ is an unentangled measurement element operator, is it possible to obtain an entangled state between A and C if Bob conducts a second measurement $\mathcal{M}_2$ on the (2,3) pair after post-processing the first measurement? To address this question, we propose two lemmas for arbitrary measurements:

\textit{Lemma 1.} If the first measurement $\mathcal{M}_1$ is a projective or rank-1 POVM (pure POVM) measurement, meaning each $\Pi_n$ element is rank 1 or projective, then any second measurement $\mathcal{M}_2$ cannot disturb the post-measurement state of $\mathcal{M}_1$ between Alice and Charlie, whether by entangling, disentangling, or increasing the entanglement.

\textit{Proof:} A POVM, $\mathcal{M}$ is called as rank 1 when all its elements are rank-1 operators. Suppose $\mathcal{M}_1$ is a projective or rank-1 POVM, implying each element $\Pi_n$ of $\mathcal{M}_1$ measurement is rank-1, such that for $\alpha=1$, $\Pi_n=\pi_{n1}|\phi_{n1}\rangle\langle\phi_{n1}|$. Then, according to Eq.\ref{eq5}, $|\Phi_n\rangle$ can be expressed as:
\begin{eqnarray}
\label{eq19}
|\Phi_n\rangle=\frac{1}{\sqrt{\Pi_n}}\sqrt{\pi_{n1}}|\phi_{n1}\rangle_{23}\otimes|\phi_{n1}^*\rangle_{14}
\end{eqnarray}
This state is separable in the $23|14$ bipartition as it can be represented as a simple tensor product in this partition. Now, if Bob attempts a measurement on (2,3) of $|\Phi_n\rangle$ to disturb the (1,4) pair, it will not be possible as $|\Phi_n\rangle$ is already separable in the $23|14$ bipartition after the first measurement. Therefore, no further measurement can disturb the (1,4) pair of the $|\Phi_n\rangle$ state.

Hence, if $\Pi_n$ is a projective or rank-1 entangled POVM element operator, the post-measurement state between A and C will be entangled, but no subsequent measurement can alter the entanglement in the (1,4) pair. Similarly, if $\Pi_n$ is a rank-1 unentangled POVM element or projective element, the post-measurement state between (1,4) will be separable, and no further sequential measurement on the (2,3) pair can entangle the (1,4) pair.

The above proof indicates that if Bob performs the first measurement with a rank greater than one, there is a possibility to disturb the (1,4) pair using another round of measurement on the (2,3) pair. This is because, from Eq.\ref{eq5}, it can be understood that the $14|23$ bipartition may exhibit non-zero entanglement after Bob's first measurement if its rank is greater than one. Consequently, one might pursue a subsequent measurement on the (2,3) pair to disturb the (1,4) pair.

 \textit{Lemma-2:} After performing the first measurement $\mathcal{M}_1$ on the (2,3) pair of the initial state Eq.\ref{eq1}, if the bipartition entanglement $\mathcal{C}_{n|14vs23}$ of the four-qubit state $|\Phi_n\rangle$ (Eq.\ref{eq5}) exhibits non-zero entanglement for a given outcome, i.e., if the $14|23$ bipartition entanglement lies within the range $0<\mathcal{C}_{n|14vs23}(|\Phi_n\rangle\langle\Phi_n|)\leq1$, and the rank of the first measurement element is greater than one, then the second measurement $\mathcal{M}_2$ has the potential to disturb the entanglement of the (1,4) pair

\textit{Proof:} In the first lemma, we have established that if $\mathcal{C}_{n|14vs23}(|\Phi_n\rangle\langle\Phi_n|)=0$, indicating zero entanglement in the $14|23$ bipartition of the post-measurement four-qubit state (Eq.\ref{eq5}), then no further measurement can disturb the entanglement of the (1,4) pair. Here, $\mathcal{C}_{n|14vs23}(|\Phi_n\rangle\langle\Phi_n|)$ represents the I-concurrence in the $14|23$ bipartition of Eq.\ref{eq5}. Using Eq.\ref{eq18}, $\mathcal{C}_{n|14vs23}$ can be expressed as:
 \begin{eqnarray}
     \label{eq20}
     \mathcal{C}_{n|14vs23}(|\Phi_n\rangle\langle\Phi_n|)&=&\sqrt{\frac{4}{3}\left[1-\mbox{Tr}(\rho_{14|n}^2)\right]}\nonumber\\
     &=&\sqrt{\frac{4}{3}\left[1-\frac{\sum_{\alpha=1}^4\pi_{n\alpha}^2}{(\mbox{Tr}\Pi_n)^2}\right]}
 \end{eqnarray}
To disturb the post-measurement state of the (1,4) pair in the next round measurement $\mathcal{M}_2$, a higher rank measurement (greater than rank one) $\mathcal{M}_1$ is required, and the post-measurement state (Eq.\ref{eq5}) must exhibit non-zero entanglement in the $14|23$ bipartition. This condition is necessary for the disturbance of the (1,4) pair in the second measurement.

In the initial state (Eq.\ref{eq1}), when no measurement was performed on the (2,3) pair, the $14|23$ bipartition entanglement was $\mathcal{C}_{14vs23}(|\Phi\rangle\langle\Phi|)=1$. However, after a particular measurement on the (2,3) pair of Eq.\ref{eq1}, one may obtain $\mathcal{C}_{n|14vs23}(|\Phi_n\rangle\langle\Phi_n|)\neq0$ or $\mathcal{C}_{n|14vs23}(|\Phi_n\rangle\langle\Phi_n|)=1$ for an outcome. This implies that the (2,3) pair is correlated with the (1,4) pair non-maximally or maximally, respectively. Therefore, any suitable further measurement on the (2,3) pair of $|\Phi_n\rangle$ (Eq.\ref{eq5}) can disturb the (1,4) pair. We will provide some examples later.

Thus far, we establish that for the second measurement $\mathcal{M}_2$ to disturb the (1,4) pair in the post-measurement state $|\Phi_n\rangle$, two conditions must be met: Firstly, the rank of the first measurement $\mathcal{M}_1$ should exceed one, and secondly, $|\Phi_n\rangle$ should exhibit non-zero entanglement in the $14|23$ bipartition (Eq.\ref{eq20}). This prompts the following question: What are the properties of measurements that can lead to entangled states in (1,4) through successive measurements if the first attempt fails? To address this inquiry, we present protocols for different categories of measurements in the entanglement swapping scenario

Initially, in the initial state (Eq.\ref{eq1})  $12|34$ bipartition entanglement is zero i.e. $\mathcal{C}_{12vs34}(|\Phi\rangle\langle\Phi|)=0$ and in the $14|23$ bipartition  $\mathcal{C}_{14vs23}(|\Phi\rangle\langle\Phi|)=1$. To get (1,4) pair entangled we need a measurement $\mathcal{M}_1$ that can produce non-zero entanglement in $12|34$ bipartition.  So, until and unless Bob's first measurement $\mathcal{M}_1$ produces entanglement in the $12|34$ bipartition we can not disturb the state between (1,4) pair during the second measurement even if $14|23$ bipartition has non-zero entanglement. Therefore, $\mathcal{C}_{n|12vs34}(|\Phi\rangle\langle\Phi|)>0$ is necessary for any type of measurement. However, this condition alone is not sufficient to entangle the (1,4) pair. There exist unentangled measurements capable of generating non-zero entanglement in the $12|34$ bipartition but unable to establish entanglement between qubits A and C. As previously established, if $\Pi_n$ is an unentangled element operator, the post-measurement state between the (1,4) pair will also be unentangled.  

To get an entangled state between A and C, we have now given the protocols for the full two-qubit measurements of different measurement categories.

\subsection{Inseparable operations}

An inseparable operation is defined as follows: let a quantum operation $\mathcal{E}_A$ acting on a system $\rho_A \in \mathcal{H}_A$ (Hilbert space), the Choi state is defined as \cite{choi}:

\begin{equation}
Choi(\mathcal{E}_A) = (\mathbb{I}_{A_1} \otimes \mathcal{E}_A) |\phi^+\rangle\langle\phi^+|_{A_1 A}
\end{equation}

Here, $\mathbb{I}_{A_1}$ is the identity operator in $\mathcal{H}_{A_1}$ with dimension $\dim(\mathcal{H}_{A_1}) = \dim(\mathcal{H}_{A}) = d$. The state $|\phi^+\rangle$ is a normalized maximally entangled state in $\mathcal{H}_{A_1} \otimes \mathcal{H}_{A}$, given by $|\phi^+\rangle = \frac{1}{\sqrt{d}} \sum_{i=0}^{d-1} |i\rangle_{A_1} |i\rangle_{A}$ where $|i\rangle_{A_1}$ and $|i\rangle_{A}$ are orthonormal bases in $\mathcal{H}_{A_1}$ and $\mathcal{H}_A$, respectively.

Consider a quantum operation $\mathcal{M}_{AB}$ acting on a bipartite state $\rho_{AB} \in \mathcal{H}_A \otimes \mathcal{H}_B$. One can relabel the bases of $\mathcal{H}_A \otimes \mathcal{H}_B$ such that it becomes $\mathcal{H}^{d^A\times d^B} = \mathcal{H}_C$. Now, the whole operation can be viewed as a single-partite operation on $\mathcal{H}_C$, and one can represent it as $(\mathbb{I}_{C_1} \otimes \mathcal{M}_C)|\phi^+\rangle\langle\phi^+|_{C_1 C}$, where $|\phi^+\rangle_{C_1 C}$ is the maximally entangled state in $\mathcal{H}_{C_1} \otimes \mathcal{H}_C$.

Alternatively, one can form two pairs of maximally entangled states of two qubits $|\phi^+\rangle\langle\phi^+|_{A_1 A}$ and $|\phi^+\rangle\langle\phi^+|_{B B_1}$ and apply the quantum operation on their tensor product:

\begin{eqnarray}
\label{eq21}
\rho_{A_1ABB_1} = (\mathbb{I}_{A_1} \otimes \mathcal{M}_{AB} \otimes \mathbb{I}_{B_1})|\phi^+\rangle \langle\phi^+|_{A_1A} \otimes |\phi^+\rangle \langle\phi^+|_{BB_1}
\end{eqnarray}

If the entanglement of the state $\rho_{A_1ABB_1}$ across the bipartition $A_1A\ |\ BB_1$ is non-zero, then the operation is called an inseparable operation; otherwise, it is a separable operation \cite{cirac}.

Both approaches are equivalent to define separable and inseparable quantum operations, as $|\phi^+\rangle\langle\phi^+|_{A_1 A} \otimes |\phi^+\rangle\langle\phi^+|_{B B_1}=|\phi^+\rangle\langle\phi^+|_{CC_1}$. We follow the definition in Eq.\ref{eq21} to differentiate the separable and inseparable operations throughout the work.


In our work, we consider two initial states, $|\phi^+\rangle$, which are maximally entangled two-qubit states. The entanglement of the four-qubit post-measurement state, $|\Phi_n\rangle$, in the $12|34$ bipartition, i.e., $\mathcal{C}_{n|12vs34}(|\Phi_n\rangle\langle\Phi_n|)$, is the inseparability condition of the $\Pi_n$ element of any arbitrary two-qubit generalized measurement from an operational point of view. We find out the exact expression of inseparability criteria \cite{cirac} for two-qubit arbitrary measurement using the I-concurrence relation \cite{rungta} which is given by (for details see Appendix \ref{ap2}):
\begin{widetext}
\begin{eqnarray}
    \label{eq22}
    \mathcal{C}_{n|12vs34}(|\Phi_n\rangle\langle\Phi_n|)&=&\sqrt{\frac{4}{3}\left[1-\mbox{Tr}(\rho_{12|n}^2)\right]}\nonumber\\
    &=&\sqrt{\frac{4}{3}\left[1-\frac{1}{(\mbox{Tr}\Pi_n)^2}\sum_{\substack{\alpha,\beta,\\ \gamma,\eta=1}}^4\sum_{\substack{i,j,i^{\prime},k,\\l,k^{\prime}=0}}^1\sqrt{\pi_{n\alpha}}\sqrt{\pi_{n\beta}}\sqrt{\pi_{n\gamma}}\sqrt{\pi_{n\eta}}a^*_{n\alpha|ij}a_{n\beta|i^{\prime}j}a^*_{n\beta|k^{\prime}l}a_{n\alpha|kl}a^*_{n\gamma|i^{\prime}j}a_{n\eta|ij}a^*_{n\eta|kl}a_{n\gamma|k^{\prime}l}\right]}.
\end{eqnarray}
\end{widetext}
Now, we present protocols for performing two-qubit measurements of different categories to obtain an entangled state between (1,4) pair when the measurement is an inseparable operation.

\subsubsection{Entangled measurements}

If at least one element $\Pi_n$ of measurement $\mathcal{M}_1$ is entangled, then the measurement is termed as an entangled measurement, which consistently yields non-zero entanglement in the $12|34$ bipartition, i.e., $\mathcal{C}_{n|12vs34}(|\Phi_n\rangle\langle\Phi_n|)>0$ (Eq.\ref{eq22}). This occurs because when an entangled POVM element $\Pi_n$ is applied to the (2,3) pair of the initial state, the resulting post-measurement state between A and C is $\frac{\Pi_n^*}{\mbox{Tr}(\Pi_n)}$, which remains entangled. Therefore, any entangled measurement constitutes an inseparable operation, ensuring non-zero entanglement in the $12|34$ bipartition of the post-measurement state for at least one outcome.

If the first measurement is entangled, the post-measurement state between the (1,4) pair is always entangled. However, whether a second measurement on (2,3) can disturb the entanglement of the (1,4) pair depends on the $14|23$ bipartition entanglement (Eq.\ref{eq20}) of the four-qubit post-measurement state resulting from the first measurement $\mathcal{M}_1$.  So, there are two types of entangled measurements for which two situations may arise:

\paragraph{\textit{$\mathcal{C}_{n|14vs23}(|\Phi_n\rangle\langle\Phi_n|)=0$}}
After the first entangled measurement, if $\mathcal{C}_{n|14vs23}(|\Phi_n\rangle\langle\Phi_n|)=0$ (Eq.\ref{eq20}) for all elements of $\mathcal{M}_1$, then further second measurement $\mathcal{M}_2$ cannot disturb the (1,4) pair anymore.

For instance, if Bob conducts a projective non-maximally entangled basis measurement, or  a Bell basis measurement, or any rank-1 entangled POVM on (2,3) of the initial state Eq.\ref{eq1}, then the resulting post-measurement state (Eq.\ref{eq6}) between A and C will be an entangled state. However, no further measurement on the (2,3) pair of Eq.\ref{eq5} can enhance, disentangle, or decrease the entanglement in the (1,4) pair. 

\paragraph{\textit{$\mathcal{C}_{n|14vs23}(|\Phi_n\rangle\langle\Phi_n|)>0$}}

If the four-qubit post-measurement state has non-zero entanglement in the $14|23$ bipartition for at least one outcome, i.e., $0<\mathcal{C}_{n|14vs23}(|\Phi_n\rangle\langle\Phi_n|)\leq1$ (Eq.\ref{eq20}), after the first measurement $\mathcal{M}_1$, whose rank is greater than one, then there would be a second measurement that could disturb the (1,4) pair again. Thus, there is a possibility to enhance, decrease, or disentangle the entanglement of the (1,4) pair. Examples are provided in a later section.

\subsubsection{Unentangled measurements}

An unentangled measurement is considered an inseparable operation in an operational sense if $\mathcal{C}_{n|12vs34}>0$ (as indicated in Eq.\ref{eq22}) for at least one element of a measurement. However, from the perspective of measurement entanglement, if each element $\Pi_n$ of measurement $\mathcal{M}_1$ is unentangled, meaning that the corresponding operator defined as $\frac{\Pi{n}}{\mbox{Tr}(\Pi_n)}$ is a separable state \cite{vertesi,adam,yokoyama,sperling,hamamura}, then the post-measurement state between A and C, i.e., the (1,4) pair, is always separable.
 However, an unentangled measurement element with non-zero entanglement in the $14|23$ bipartition of the post-measurement state can still create entanglement between the (1,4) pair if it conforms to non-zero entanglement in the $12|34$ bipartition and another suitable measurement $\mathcal{M}_2$ is done on the (2,3) pair of the post-processed four qubits state. That means subsequent measurement can disturb the (1,4) pair only if the first measurement is an inseparable operation and retains $14|23$ bipartition entanglement in the post-measurement state. Two different circumstances are observed based on the $14|23$ bipartition entanglement of the post-measurement state:

\paragraph{\textit{$\mathcal{C}_{n|14vs23}(|\Phi_n\rangle\langle\Phi_n|)=0$}}

It is not possible to find a measurement that can create non-zero entanglement in the $12|34$ bipartition of the post-measurement state, i.e., $\mathcal{C}_{n|12vs34}(|\Phi_n\rangle\langle\Phi_n|)>0$, and $\mathcal{C}_{n|14vs23}(|\Phi_n\rangle\langle\Phi_n|)=0$  for an element. $\mathcal{C}_{n|14vs23}=\sqrt{\frac{4}{3}[1-\mbox{Tr}(\rho_{14|n}^2)]}=0$ implies that $\mbox{Tr}(\rho_{14|n}^2)=1$, indicating that the post-measurement states $\rho_{14|n}$  must be pure, suggesting the presence of a projective or rank-1 POVM element. For an unentangled rank-1 POVM element or an unentangled projective element, $\Pi_n$ must be in the form of a tensor product, thereby preventing the generation of non-zero entanglement in the $12|34$ bipartition of $|\Phi_n\rangle$.

\paragraph{\textit{$\mathcal{C}_{n|14vs23}(|\Phi_n\rangle\langle\Phi_n|)>0$}} 

If the first measurement $\mathcal{M}_1$ results in non-zero entanglement in the $12|34$ bipartition of the post-measurement state (Eq. \ref{eq5}) for one outcome (i.e., $\mathcal{C}_{n|12vs34}(|\Phi_n\rangle\langle\Phi_n|)>0$), and a second measurement $\mathcal{M}_2$ acts on the (2,3) pair successively, the final post-measurement state (Eq.\ref{eq17}) between A and C will become entangled with a suitable choice of measurements. We determine the entanglement criteria of the final post-measurement state $\rho_{14|nm}$ (Eq.\ref{eq17}) between A and C in terms of measurement operators. We derive the exact expression of the negativity \cite{vidal} of the post-measurement state of the (1,4) pair after the sequential measurements in Appendix \ref{ap3}. To evaluate the entanglement of the final state $\rho_{14|nm}$, we use the negativity measure \cite{vidal} to accommodate the arbitrariness of the measurements. We do not utilize the concurrence measure due to the difficulty in obtaining an expression for an arbitrary state without further information about the state's parameters.

When Bob performs an unentangled measurement, $\mathcal{M}_1$, which is unable to establish entanglement between the (1,4) pair, the entanglement in the $14|23$ bipartition is retained in the post-measurement state. A second measurement, $\mathcal{M}_2$, can then establish entanglement in the final post-measurement state, $\rho_{14|nm}$, if the negativity, $\mathcal{N}(\rho_{14|nm})$, is greater than zero.

As an example, we consider a Bell measurement with white noise as the first measurement $\mathcal{M}_1$. The POVM elements are defined as 
\begin{eqnarray}
    \label{eq29}
    \Pi_1&=&\lambda|\phi^+\rangle\langle\phi^+|+\frac{1-\lambda}{4}\mathbb{I}_4\nonumber\\
    &=&\frac{3\lambda+1}{4}|\phi^+\rangle\langle\phi^+|+\frac{1-\lambda}{4}\Big(|\phi^-\rangle\langle\phi^-|+|\psi^+\rangle\langle\psi^+|+|\psi^-\rangle\langle\psi^-|\Big)\nonumber\\
    \Pi_2&=&\lambda|\phi^-\rangle\langle\phi^-|+\frac{1-\lambda}{4}\mathbb{I}_4\nonumber\\
    &=&\frac{3\lambda+1}{4}|\phi^-\rangle\langle\phi^-|+\frac{1-\lambda}{4}\Big(|\phi^+\rangle\langle\phi^+|+|\psi^+\rangle\langle\psi^+|+|\psi^-\rangle\langle\psi^-|\Big)\nonumber\\
    \Pi_3&=&\lambda|\psi^+\rangle\langle\psi^+|+\frac{1-\lambda}{4}\mathbb{I}_4\nonumber\\
    &=&\frac{3\lambda+1}{4}|\psi^+\rangle\langle\psi^+|+\frac{1-\lambda}{4}\Big(|\phi^-\rangle\langle\phi^-|+|\phi^+\rangle\langle\phi^+|+|\psi^-\rangle\langle\psi^-|\Big)\nonumber\\
    \Pi_4&=&\lambda|\psi^-\rangle\langle\psi^-|+\frac{1-\lambda}{4}\mathbb{I}_4\nonumber\\
    &=&\frac{3\lambda+1}{4}|\psi^-\rangle\langle\psi^-|+\frac{1-\lambda}{4}\Big(|\phi^-\rangle\langle\phi^-|+|\psi^+\rangle\langle\psi^+|+|\phi^+\rangle\langle\phi^+|\Big)\nonumber\\
\end{eqnarray}
for $\lambda\in[0,1]$. The measurement is unentangled  for $0\leq\lambda\leq\frac{1}{3}$ and entangled for $\frac{1}{3}<\lambda\leq1$. The rank of the each element $\Pi_n$ is four.  Then from Eq.\ref{eq4} we find that $p_n=\frac{1}{4}$, all outcomes are equally probable. 

For the outcome $n$, the post-measurement states between A and C are obtained
from Eq.\ref{eq6}; in particular,
\begin{eqnarray}
    \label{eq30}
    \rho_{14|n}=\Pi_n(\lambda)
\end{eqnarray}
for $n=1,2,3,4$. The amount of negativity of $\rho_{14|n}$ for $n\in\{1,2,3,4\}$ is 
\begin{equation}
    \mathcal{N}(\rho_{14|n})=\frac{3\lambda-1}{2}.
\end{equation}

Note that, the state $\rho_{14|n}$ is separable for $0\leq\lambda\leq\frac{1}{3}$ and entangled for $\frac{1}{3}<\lambda\leq1$.  Hence, it is reflected that the state $\rho_{14|n}$ is entangled if and only if the measurement element is entangled.
 So, the average negativity after completion of measurement $\mathcal{M}_1$  between A and C is
\begin{eqnarray}
    \label{eq31}
    \mathcal{N}_{avg}(\rho_{14|\mathcal{M}_1})&=&\sum_{n=1}^4p_n\mathcal{N}(\rho_{14|n})
    =\frac{3\lambda-1}{2}
\end{eqnarray}

Now using Eq.\ref{eq22}, we obtain the concurrence of $12|34$ bipartition of post-measurement state $|\Phi_n\rangle$ (Eq.\ref{eq5}) for each POVM elements which is given by
\begin{eqnarray}
    \label{eq32}
    \mathcal{C}_{n|12vs34}= \frac{\sqrt{1+\lambda^2-\sqrt{1-\lambda}\sqrt{1+3\lambda}+\lambda\sqrt{1-\lambda}\sqrt{1+3\lambda}}}{\sqrt{2}}.\nonumber\\
\end{eqnarray}
$\forall n\in\{1,2,\ldots,4\}$.
From the above Eq.\ref{eq32}, we observe that each element $\Pi_n$ exhibits non-zero entanglement in the $12|34$ bipartition of the post-measurement state (Eq.\ref{eq5}) for $0 < \lambda \leq 1$. Even when each $\Pi_n$ is unentangled for $0 \leq \lambda \leq \frac{1}{3}$, it still demonstrates non-zero entanglement in the $12|34$ bipartition of the four-qubit post-measurement state (Eq.\ref{eq5}), except at $\lambda = 0$.

In order to determine whether the second measurement $\mathcal{M}_2$ can disturb the post-measurement state $\rho_{14|n}$ of the first measurement $\mathcal{M}_1$ or not, we evaluate the concurrence of $14|23$ bipartition of the post-measurement state of each element $\Pi_n$. Using Eq.\ref{eq20} we obtain the concurrence of $14|23$ bipartition entanglement of the post-measurement state for each element $\Pi_n$ as:
\begin{eqnarray}
    \label{eq33}
    \mathcal{C}_{n|14vs23}=\sqrt{1-\lambda^2}
\end{eqnarray}
which is non-zero for $0\leq\lambda<1$ except at $\lambda=1$.
When $\lambda=1$, the measurement becomes a perfect Bell basis measurement, which is a projective measurement. If a Bell measurement is performed initially on the (2,3) pair, no subsequent measurements can affect the post-measurement state $\rho_{14|n}$. However, when the measurement parameter $\lambda$ is between $0\leq\lambda<1$, a subsequent measurement can disturb the post-measurement state of the first measurement.

Now, after post-processing each measurement outcome of the first measurement (Eq.\ref{eq29}), Bob performs another measurement $\mathcal{M}_2$ on the (2,3) pair. We consider the second measurement $\mathcal{M}_2$ to be a purely separable operation. The POVM elements of $\mathcal{M}_2$ are defined as:
\begin{eqnarray}
    \label{eq34}
    E_1=|\psi_{11}\rangle\langle\psi_{11}|+|\psi_{12}\rangle\langle\psi_{12}|\nonumber\\
    E_2=|\psi_{21}\rangle\langle\psi_{21}|+|\psi_{22}\rangle\langle\psi_{22}|
\end{eqnarray}
where $|\psi_{11}\rangle=|00\rangle$, $|\psi_{12}\rangle=|01\rangle$, $|\psi_{21}\rangle=|10\rangle$, and $|\psi_{22}\rangle=|11\rangle$. This measurement is a single qubit measurement performed on qubit 2 only, and does not affect qubit 3.  Each element $E_m$ is unentangled and does not have any capacity to produce an entangled state as it is a purely separable operation \cite{cirac}.  

Then, from Eq.\ref{eq15}, we find that $s_{nm}=\frac{1}{2}$, which is the probability of obtaining the $m$-th outcome of the second measurement, $\mathcal{M}_2$, when it acts on the $n$-th outcome of the first measurement, $\mathcal{M}_1$. The probability of obtaining the outcome state $\rho_{14|nm}$ between A and C is $p_n s_{nm}=\frac{1}{8}$, where all outcomes are equally probable. The post-measurement states between A and C are obtained following Eq.\ref{eq17}
\begin{eqnarray}
    \label{eq35}
    \rho_{14|11}=\frac{(1+\lambda)}{2}|\xi_{11}\rangle\langle\xi_{11}|+\frac{(1-\lambda)}{2}|01\rangle\langle01|
\end{eqnarray}
where $|\xi_{11}\rangle=\frac{\sqrt{1+3\lambda}+\sqrt{1-\lambda}}{2\sqrt{1+\lambda}}|00\rangle+\frac{\sqrt{1+3\lambda}-\sqrt{1-\lambda}}{2\sqrt{1+\lambda}}|11\rangle$. Similarly, $\rho_{14|nm}$ for other values of $n$ and $m$ can be calculated easily  by following the similar method. 

 We get the amount of negativity of the state $\rho_{14|nm}$
which is given by 
\begin{eqnarray}
    \label{eq36}
    \mathcal{N}(\rho_{14|nm})=\frac{\lambda-1+\sqrt{1-2\lambda+5\lambda^2}}{2}\nonumber\\
    \, \, \, \forall n \in \{1,2,3,4\} \, \, \, \text{and} \, \, \, \forall m \in \{1,2\}
\end{eqnarray}
We observe that $\mathcal{N}(\rho_{14|nm})$ is non-zero for the entire range of $\lambda$, except for $\lambda=0$. On the other hand, after the first measurement $\mathcal{M}_1$ alone, $\mathcal{N}(\rho_{14|n})$ is zero for $0\leq\lambda\leq\frac{1}{3}$ as the measurement is unentangled in that range, and non-zero for $\frac{1}{3}<\lambda\leq1$ as the measurement is entangled in that range. However, two successive measurements establish entanglement between A and C for the entire range of the measurement parameter $\lambda$, except for $\lambda=0$. Thus, successive measurements open up the entanglement between A and C for the entire range of $\lambda$. 

For every outcome, the amount of negativity is same. The average negativity between A and C after the completion of two measurements $\mathcal{M}_1$ and $\mathcal{M}_2$ in succession is 
\begin{eqnarray}
    \label{eq37}
    \mathcal{N}_{avg}(\rho_{14|\mathcal{M}_1\mathcal{M}_2})&=&\sum_{n=1}^4\sum_{m=1}^2p_ns_{nm}\mathcal{N}(\rho_{14|nm})\nonumber\\
    &=&\frac{\lambda-1+\sqrt{1-2\lambda+5\lambda^2}}{2}
\end{eqnarray}

\begin{figure}[h!]
 \includegraphics[width=7cm, height=5cm]{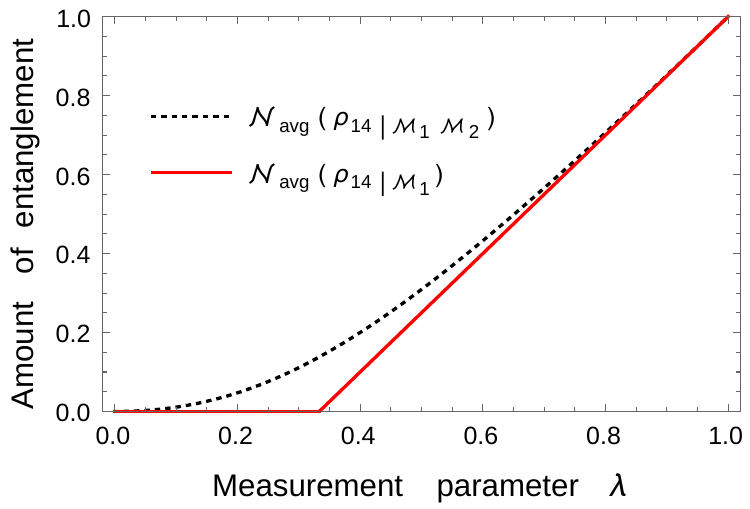}
 \caption{Variation of negativity with the measurement parameter $\lambda$ for the average entanglement between A and C. Here $\mathcal{N}_{avg}(\rho_{14|\mathcal{M}_1\mathcal{M}_2}) $ represents the average negativity established between A and C after the completion of two measurements $\mathcal{M}_1$ and $\mathcal{M}_2$ in succession. $\mathcal{N}_{avg}(\rho_{14|\mathcal{M}_1})$ represents the average negativity between A and C after the completion of $\mathcal{M}_1$ measurement.}
 \label{fig1}
\end{figure}
In Fig.\ref{fig1}, we have plotted the average entanglement (Eq.\ref{eq37} and Eq.\ref{eq31}) established between A and C with respect to the measurement parameter $\lambda$ after the completion of the successive two measurements  $\mathcal{M}_1$ followed by $\mathcal{M}_2$ and after the completion of the first measurement $\mathcal{M}_1$ in terms of negativity measure.    We make the following observations: (i) After the first measurement $\mathcal{M}_1$ the average negativity $\mathcal{N}_{avg}(\rho_{14|\mathcal{M}_1})$ established between A and C is zero in between $0\leq\lambda\leq\frac{1}{3}$ because the measurement is not entangled in that range. (ii) Due to two successive measurements the average negativity $\mathcal{N}_{avg}(\rho_{14|\mathcal{M}_1\mathcal{M}_2})$ is non-zero for the whole range of the $0<\lambda\leq1$ and increasing with the $\lambda$ where $\mathcal{M}_2$ is a separable operation.  (iii) The average negativity $\mathcal{N}_{avg}(\rho_{14|\mathcal{M}_1\mathcal{M}_2})$ established between A and C after the two successive measurements is greater than the first measurement for the whole range of the $\lambda$.

An unentangled measurement $\mathcal{M}_1$ (Eq. \ref{eq29}) within the range $0\leq\lambda\leq\frac{1}{3}$, along with a purely separable operation $\mathcal{M}_2$ (Eq. \ref{eq34}), individually falls short in establishing entanglement between parties A and C. However, when executed sequentially, with $\mathcal{M}_1$ followed by $\mathcal{M}_2$, entanglement between A and C emerges across the entire range of $\lambda$, except for $\lambda=0$. Furthermore, this sequential approach yields a greater amount of entanglement compared to a single measurement $\mathcal{M}_1$, particularly in terms of the entanglement established between A and C.

Meanwhile, an unentangled measurement, characterized by $\mathcal{C}_{n|12vs34}=0$ (Eq. \ref{eq22}) for every outcome, proves ineffective in fostering entanglement between A and C, as previously elaborated. In the upcoming section, we delve into the prospect of attaining entanglement between A and C through a subsequent measurement following Bob's purely separable operation.

\subsection{Separable operations}

A measurement that operates as a separable operation is incapable of transforming separable states into entangled ones \cite{rain1}. Therefore, if Bob initiates a measurement that constitutes a separable operation, it will not induce entanglement in the $12|34$ bipartition of the post-measurement state, signifying $\mathcal{C}_{n|12vs34}=0$ (Eq.\ref{eq22}) for each POVM element. This implies that the post-measurement states of the (1,2) and (3,4) pairs must be pure for every outcome, as $\mathcal{C}_{n|12vs34}=0$ implies $\mbox{Tr}(\rho_{12|n}^2)=1$ and $\mbox{Tr}(\rho_{34|n}^2)=1$. Thus, in our scenario, the post-measurement states within the (1,2) and (3,4) pairs remain pure. 

Can a second measurement entangle the (1,4) pair after Bob conducts a separable operation? We affirmatively address this question.

A separable operation $\{\Pi_n\}$ can be represented as:
\begin{equation}
\label{eq38}
\Pi_n=\mathcal{A}_n\otimes\mathcal{B}_n
\end{equation}
where $\sum_n\Pi_n=\mathbb{I}$ and $\mathcal{A}_n$, $\mathcal{B}_n$ are positive operators acting on $\mathcal{H}_{\mathcal{A}}$, $\mathcal{H}_{\mathcal{B}}$ Hilbert spaces respectively. In our scenario, if Bob conducts a separable operation on the (2,3) pair and the operators $\mathcal{A}_n$ and $\mathcal{B}_n$ are rank-1 for each element, there will be no second measurement $\mathcal{M}_2$ capable of disturbing the (1,4) pair. This is because a rank-one element on each qubit destroys the initial entanglement of $|\phi^+\rangle_{12}$ and $|\phi^+\rangle_{34}$ states.

However, if $\mathcal{A}_n$ and $\mathcal{B}_n$ are rank-2 elements, it may be possible to retain entanglement in the post-measurement states of the (1,2) and (3,4) pairs. In such a case, during the next round of measurement, there may be an opportunity to entangle the (1,4) pair by selecting a suitable measurement after post-processing the first measurement $\mathcal{M}_1$. Here, the second measurement $\mathcal{M}_2$ should be an inseparable operation, i.e., it should have $\mathcal{C}_{n|12vs34}>0$ (Eq.\ref{eq22}) for at least one element. It's important to note that to achieve an entangled state in the (1,4) pair using two successive measurements, Eq.\ref{eq28} must be non-zero for an outcome

Let's delve into the evaluation of the $\mathcal{A}_n$ and $\mathcal{B}_n$ elements that preserve the entanglement in the (1,2) and (3,4) pairs following the first measurement $\mathcal{M}_1$, thus allowing the potential entanglement of the (1,4) pair through a suitable second measurement.

Suppose a single-qubit measurement element, $\mathcal{A}_n$, acts on qubit 2 of the initial state $|\Psi\rangle_{12}=|\phi^+\rangle_{12}=\frac{1}{\sqrt{2}}(|00\rangle+|11\rangle)$. Similarly, $\mathcal{B}_n$ acts on qubit 3 of the initial state $|\Psi\rangle_{34}=|\phi^+\rangle_{34}$. We assess the post-measurement state for $\mathcal{A}_n$ while retaining the arbitrary nature of the measurement. The same methodology applies to $\mathcal{B}_n$ also.

The $\mathcal{A}_n$ element can be decomposed as 
$\mathcal{A}_n=\sum_{\alpha=1}^2\tau_{n\alpha}|\varphi_{n\alpha}\rangle\langle\varphi_{n\alpha}|$ where $|\varphi_{n\alpha}\rangle$, $\alpha=1,2$ form a complete orthonormal basis on $\mathbb{C}^2$,  and $\tau_{n\alpha}\geq0$ for $\alpha=1,2$.  If $\mathcal{A}_n$ acts on qubit 2 of $|\phi^+\rangle_{12}$, then the post-measurement state becomes:
\begin{eqnarray}
    \label{eq39}
    |\Psi^{\prime}\rangle_n&=& \mathbb{I}_2\otimes\sqrt{\mathcal{A}_n}|\Psi\rangle_{12}\nonumber\\
   &=&\sum_{i,j=0}^1\sum_{\alpha=1}^2a_{ij}\sqrt{\tau_{n\alpha}}\langle\varphi_{n\alpha}|j\rangle|i\rangle_1\otimes|\varphi_{n\alpha}\rangle_2
\end{eqnarray}
which remains a pure state.

From Eq.\ref{eq39}, it is evident that when the measurement element is a rank-1 POVM or projective, the post-measurement state is always separable. However, when the measurement element's rank is greater than one, there is a possibility that the post-measurement state remains entangled. By considering arbitrary orthonormal bases for each element: $|\varphi_{n1}\rangle=\cos\frac{\theta_n}{2}|0\rangle+e^{i\phi_n}\sin\frac{\theta_n}{2}|1\rangle$ and $|\varphi_{n2}\rangle=\sin\frac{\theta_n}{2}|0\rangle-e^{i\phi_n}\cos\frac{\theta_n}{2}|1\rangle$ on $\mathbb{C}^2$, we evaluate the I-concurrence \cite{rungta} of the post-measurement state of each outcome using Eq.\ref{eq18}:
\begin{eqnarray}
\label{eq40}
\mathcal{C}_n(|\Psi^{\prime}\rangle_n\langle\Psi^{\prime}|)=\frac{2\sqrt{\tau_{n1}\tau_{n2}}}{\tau_{n1}+\tau_{n2}}
\end{eqnarray}
where we assume $|\Psi\rangle_{12}=|\phi^+\rangle$, a Bell state. Eqquation (\ref{eq40}) reveals that any rank-2 measurement element retains the entanglement in the (1,2) and (3,4) pairs. Therefore, in the second round of measurement, if Bob conducts an inseparable operation on the (2,3) pair after post-processing the first measurement $\mathcal{M}_1$, there's a possibility of obtaining an entangled state in the (1,4) pair. However, it's crucial to note that the second measurement should be an inseparable operation, ensuring that the condition $\mathcal{C}_{n|12vs34}>0$ for an outcome [as stated in Eq. \ref{eq22}] is met. 

\paragraph*{}It is noteworthy that our protocols are not only applicable for two rounds of successive measurements, but one can proceed further for $n$ rounds by following the same procedure until the (1,4) pair becomes entangled. After each round of measurement, the entanglement between the $14|23$ bipartition of the four qubits post-measurement state must be non-zero,  for an outcome, in order for the next round of measurement to disturb the (1,4) pair, after post-processing the previous round measurements. We already know that a measurement will be successful in establishing entanglement in the (1,4) pair if it can create entanglement between the $12|34$ bipartition of the total system state.

 Our protocol is also applicable when starting with two copies of $d \times d$ maximally entangled states, instead of two-qubit maximally entangled states. S. Yokoyama et al. \cite{yokoyama} have shown that by considering the $d \times d$ maximally entangled states, the post-measurement state in the (1,4) pair will be $\frac{\Pi_n^*}{\mbox{Tr}(\Pi_n)}$ if a generalized POVM element $\Pi_n$ is done on the (2,3) pair. The proof can be easily carried out by following our results and considering that every POVM element can be decomposed into $d^2$ complete orthonormal basis on $\mathbb{C}^d\otimes\mathbb{C}^d$ and $\pi_{n\alpha}\geq0$ for all $n$. To calculate $\mathcal{C}_{n|14vs23}$ for each POVM element using Eq.\ref{eq20}, one must follow the same procedure, but with the sum running from $\alpha=1$ to $\alpha=d^2$ for $d \times d$ maximally entangled states. Similarly, in the case of Eq.\ref{eq22}, one must sum for $i,j,l,k,i^\prime,k^\prime=0,...,d-1$ and for $\alpha,\beta,\gamma,\eta=1,...,d^2$. Here, Lemma-1 and Lemma-2 are also applicable in this scenario. Therefore, our protocol is also valid in this scenario, if one starts with $d \times d$ maximally entangled states. However, in this scenario, one must deal with bound entangled states and Positive Partial Transpose (PPT) measurements \cite{virmani,horodecki1,horodecki2,bennett1}, among other considerations.

\section{CONCLUSIONS}\label{sec2}

In this study, we explore entanglement activation in a generalized entanglement swapping process involving two Bell pairs and generalized measurements. Conventional wisdom suggests that entangled measurements are required to establish entanglement between distant parties.

Our work reassesses the necessity and sufficiency conditions for measurement operators to be entangled within the framework of entanglement activation in a generalized entanglement swapping process. We begin with two Bell pairs: (1, 2) shared between Alice and Bob, and (3, 4) shared between Bob and Charlie. A quantum measurement, characterized by positive operator-valued measure (POVM), is applied to qubits (2, 3), resulting in the creation of a shared state in (1, 4) between the spatially separated observers.

Our findings demonstrate that while an entangled measurement is sufficient, it is not mandatory for the measurement operators to be entangled to establish an entangled state between A and C. We introduce a sequential approach where Bob conducts an initial measurement, followed by another measurement if the initial one fails to establish the entanglement. This sequential strategy, in conjunction with appropriate post-processing after the initial measurement, facilitates the establishment of entanglement between A and C. Our investigation offers unique insights into the role of measurement operators in entanglement generation.

Our findings have identified specific criteria for different measurement operators that enable the potential for performing a second measurement to establish entanglement.  We have delineated protocols for cases where the first measurement fails to establish entanglement, showcasing the feasibility of generating entanglement between distant parties through a combination of measurements. We have found out the exact expression inseparability condition of an arbitrary two-qubit measurement.

Furthermore, our demonstration has underscored the significance of the $14|23$ bipartition entanglement of the total system state, highlighting the role of the measurement's rank in facilitating a disturbance of the (1,4) state by the second measurement. We establish that following the first measurement, the $14|23$ bipartition entanglement of the total system state must fall within the range of 0 to 1, with the measurement's rank exceeding one to enable such disturbance of the (1,4) state by the second measurement. Moreover, we have demonstrated that achieving a zero $14|23$ bipartition entanglement due to an inseparable operator in the measurement is unattainable.

We initiated our analysis with an illustration involving a measurement in the Bell basis affected by white noise, revealing its inability to establish entanglement within the range $0\leq\lambda\leq\frac{1}{3}$ due to its lack of entanglement in this interval. However, through the implementation of two consecutive measurements, we observed that the average entanglement remains non-zero across the entire range of $0<\lambda\leq1$, progressively increasing with $\lambda$, where the second measurement is executed as a separable operation. Notably, the average entanglement achieved between parties A and C following the two successive measurements surpasses that of the initial measurement across the entire $\lambda$ range. Our analysis elucidates how successive measurements outperform single measurements, offering practical benefits of entanglement distribution in quantum networks.
 
 Moreover, our approach's versatility extends beyond bipartite qubit states to higher-dimensional maximally entangled states, highlighting its applicability across various quantum scenarios.

In conclusion, expanding the protocol to include non-maximally entangled states would enhance its applicability, while extending it to cover a broader range of scenarios holds promise for further advancements in the field.
We have additionally addressed this matter by considering pure non-maximally entangled states ($\cos{\theta}|00\rangle+\sin{\theta}|11\rangle$) as initial states for the qubit pairs (1,2) and (3,4). This state exhibits entanglement for $0<\theta<\frac{\pi}{2}$, reaching maximum entanglement at $\theta=\frac{\pi}{4}$ with a concurrence of $2\cos\theta\sin\theta$. We have then performed sequential measurements using Bell basis with white noise as the first POVM measurement Eq.\ref{eq29}, followed by a second measurement Eq.\ref{eq34}. After the first measurement, the entanglement of the output state in the (1,4) pair depends on both $\theta$ and $\lambda$. However, we did not observe a state that remains entangled across the entire range of $\lambda$ for a fixed value of $\theta$. Conversely, the sequential application of these two measurements results in the (1,4) pair being entangled across the entire range of $\lambda$ for $0<\theta<\frac{\pi}{2}$. Therefore, our protocol is also applicable in scenarios involving pure non-maximally entangled states for entanglement distribution. Despite the challenges in obtaining a compact expression for the post-measurement state in (1,4) pair under arbitrary measurements with non-maximally entangled initial states, our findings suggest promising avenues for future research.

Future research efforts should prioritize refining entanglement activation protocols, exploring innovative measurement strategies, and broadening the protocol's applicability to diverse scenarios, including nonlocality activation \cite{klobus,gisin} and network nonlocality \cite{bran1,bran2}. By addressing these research directions, we can deepen our understanding of entanglement generation and distribution, leading to advancements in quantum communication and computation.

\section*{Acknowledgements} The authors express gratitude to Professor S. Bandyopadhyay for fruitful discussions. PB acknowledges S. Halder for helpful discussions.

\onecolumngrid
\appendix
\begin{center}
   {\bf \Large Appendix} 
\end{center}

\section{Post-measurement state of (1,2) and (3,4) pairs}\label{ap1}
To get the post-measurement state of (1,2) and (3,4) for $n$th outcome, we rewrite  Eq.\ref{eq5} in computational basis form. Every orthonormal basis $|\phi_{n\alpha}\rangle$ can be written in the following computational basis form $|\phi_{n\alpha}\rangle=\sum_{i,j=0}^1 a_{n\alpha|ij}|ij\rangle$ where $\sum_{i,j=0}^1|a_{n\alpha|ij}|^2=1$ and $a_{n\alpha|ij}\in\mathbb{C}$:

\begin{eqnarray}
    \label{eq7}
    |\Phi_n\rangle=\frac{1}{2\sqrt{p_n}}\sum_{\alpha=1}^4\sum_{i,j,k,l=0}^1\sqrt{\pi_{n\alpha}}a^*_{n\alpha|ij}a_{n\alpha|kl}|ij\rangle_{14}|kl\rangle_{23}
\end{eqnarray}
The density of state of Eq.\ref{eq7} is
\begin{eqnarray}
    \label{eq8}
    \rho_{1234|n}&=&|\Phi_n\rangle\langle\Phi_n|\nonumber\\
    &=&\frac{1}{4p_n}\sum_{\alpha,\beta=1}^4\sum_{\substack{i,j,k,l,\\ i^{\prime},j^{\prime},\\ k^{\prime},l^{\prime}=0}}^1\sqrt{\pi_{n\alpha}}\sqrt{\pi_{n\beta}}a^*_{n\alpha|ij}a_{n\alpha|kl}a_{n\beta|i^{\prime}j^{\prime}}a^{\ast}_{n\beta|k^{\prime}l^{\prime}} |ij\rangle\langle i^{\prime}j^{\prime}|_{14}\otimes|kl\rangle\langle k^{\prime}l^{\prime}|_{23}
\end{eqnarray}
 From the orthogonality condition of the orthonormal basis, we get 
\begin{eqnarray}
    \label{eq9}
    \langle\phi_{n\beta}|\phi_{n\alpha}\rangle&=&\sum_{i,j,k,l=0}^{1}a^*_{n\beta|kl}a_{n\alpha|ij}\langle kl|ij\rangle\nonumber\\
    &=&\delta_{\beta\alpha}
\end{eqnarray}
which reduce to
\begin{eqnarray}
    \label{eq10}
    \sum_{i,j=0}^1a^*_{n\beta|ij}a_{n\alpha|ij}=\delta_{\beta\alpha}
\end{eqnarray}

 Tracing out the three and four qubits from Eq.\ref{eq8} gives $j=j^{\prime}$ and $l=l^{\prime}$. So, we get the post-measurement state of $n$-th outcome between (1,2) pair which is given by
\begin{eqnarray}
    \label{eq11}
    \rho_{12|n}&=&\frac{1}{4p_n}\sum_{\alpha,\beta=1}^4\sum_{\substack{i,j,k,l,\\ i^{\prime}, k^{\prime}=0}}^1\sqrt{\pi_{n\alpha}}\sqrt{\pi_{n\beta}}a^*_{n\alpha|ij}a_{n\alpha|kl}a_{n\beta|i^{\prime}j}a^{\ast}_{n\beta|k^{\prime}l}|ik\rangle\langle i^{\prime}k^{\prime}|_{12}\nonumber\\
   &=&\frac{1}{\mbox{Tr}(\Pi_n)}\sum_{\alpha,\beta=1}^4\sum_{\substack{i,j,k,l,\\ i^{\prime}, k^{\prime}=0}}^1\sqrt{\pi_{n\alpha}}\sqrt{\pi_{n\beta}}a^*_{n\alpha|ij}a_{n\alpha|kl}a_{n\beta|i^{\prime}j}a^{\ast}_{n\beta|k^{\prime}l}|ik\rangle\langle i^{\prime}k^{\prime}|_{12}\nonumber\\
\end{eqnarray}
Similarly, tracing out one and two qubits from Eq.\ref{eq8} gives $i=i^{\prime}$ and $k=k^{\prime}$. The post-measurement state of (3,4) pair of $n$th outcome is

\begin{eqnarray}
    \label{eq12}
    \rho_{34|n}&=&\frac{1}{\mbox{Tr}(\Pi_n)}\sum_{\alpha,\beta=1}^4\sum_{\substack{i,j,k,l,\\ j^{\prime}, l^{\prime}=0}}^1\sqrt{\pi_{n\alpha}}\sqrt{\pi_{n\beta}}a^*_{n\alpha|ij}a_{n\alpha|kl}a_{n\beta|ij^{\prime}}a^{\ast}_{n\beta|kl^{\prime}}|lj\rangle\langle l^{\prime}j^{\prime}|_{34}
\end{eqnarray}
From the  Eq.\ref{eq11}, and Eq.\ref{eq12} one can easily compute the post-measurement states of $n$-th outcome between different pairs (1,2), and (3,4) respectively after the $\mathcal{M}_1$ POVM measurement.

\section{Inseparable quantum operation condition of an arbitrary two qubits measurement:}\label{ap2}

Amount of entanglement in $12|34$ bipartition of the post-measurement state for $n$-th outcome is given by:
\begin{eqnarray}
    \label{eqa1}
    \mathcal{C}_{n|12vs34}(|\Phi_n\rangle\langle\Phi_n|)&=&\sqrt{\frac{4}{3}\left[1-\mbox{Tr}(\rho_{12|n}^2)\right]}
\end{eqnarray}
Now, from Eq.\ref{eq11} we get that
\begin{eqnarray}
    \label{eqa2}
    \rho_{12|n}&=&\frac{1}{4p_n}\sum_{\alpha,\beta=1}^4\sum_{\substack{i,j,k,l,\\ i^{\prime}, k^{\prime}=0}}^1\sqrt{\pi_{n\alpha}}\sqrt{\pi_{n\beta}}a^*_{n\alpha|ij}a_{n\alpha|kl}a_{n\beta|i^{\prime}j}a^{\ast}_{n\beta|k^{\prime}l}|ik\rangle\langle i^{\prime}k^{\prime}|_{12}\nonumber\\
   &=&\frac{1}{\mbox{Tr}(\Pi_n)}\sum_{\alpha,\beta=1}^4\sum_{\substack{i,j,k,l,\\ i^{\prime}, k^{\prime}=0}}^1\sqrt{\pi_{n\alpha}}\sqrt{\pi_{n\beta}}a^*_{n\alpha|ij}a_{n\alpha|kl}a_{n\beta|i^{\prime}j}a^{\ast}_{n\beta|k^{\prime}l}|ik\rangle\langle i^{\prime}k^{\prime}|_{12}
\end{eqnarray}
So, $\rho_{12|n}^2$ can be written as 
\begin{eqnarray}
    \label{eqa3}
    \rho_{12|n}^2=\frac{1}{(\mbox{Tr}\Pi_n)^2}\sum_{\substack{\alpha,\beta,\\ \gamma,\eta=1}}^4\sum_{\substack{i,j,i^{\prime},k,\\l,k^{\prime},p,q,\\p^{\prime},q^{\prime}=0}}^1\sqrt{\pi_{n\alpha}}\sqrt{\pi_{n\beta}}\sqrt{\pi_{n\gamma}}\sqrt{\pi_{n\eta}}a^*_{n\alpha|ij}a_{n\beta|i^{\prime}j}a^*_{n\beta|k^{\prime}l}a_{n\alpha|kl}a^*_{n\gamma|pj}a_{n\eta|p^{\prime}j}a^*_{n\eta|q^{\prime}l}a_{n\gamma|ql}|ik\rangle\langle i^{\prime}k^{\prime}|pq\rangle\langle p^{\prime}q^{\prime}|
\end{eqnarray}
The  Eq.\ref{eqa3} can be simplified as 
\begin{eqnarray}
    \label{eqa4}
    \rho_{12|n}^2=\frac{1}{(\mbox{Tr}\Pi_n)^2}\sum_{\substack{\alpha,\beta,\\ \gamma,\eta=1}}^4\sum_{\substack{i,j,i^{\prime},k,\\l,k^{\prime},\\p^{\prime},q^{\prime}=0}}^1\sqrt{\pi_{n\alpha}}\sqrt{\pi_{n\beta}}\sqrt{\pi_{n\gamma}}\sqrt{\pi_{n\eta}}a^*_{n\alpha|ij}a_{n\beta|i^{\prime}j}a^*_{n\beta|k^{\prime}l}a_{n\alpha|kl}a^*_{n\gamma|i^{\prime}j}a_{n\eta|p^{\prime}j}a^*_{n\eta|q^{\prime}l}a_{n\gamma|k^{\prime}l}|ik\rangle\langle p^{\prime}q^{\prime}|
\end{eqnarray}
as $p=i^{\prime}$ and $k^{\prime}=q$.

So, $\mbox{Tr}(\rho_{12|n}^2)$ is given by
\begin{eqnarray}
    \label{eqa5}
    \mbox{Tr}(\rho_{12|n}^2)=\frac{1}{(\mbox{Tr}\Pi_n)^2}\sum_{\substack{\alpha,\beta,\\ \gamma,\eta=1}}^4\sum_{\substack{i,j,i^{\prime},k,\\l,k^{\prime}=0}}^1\sqrt{\pi_{n\alpha}}\sqrt{\pi_{n\beta}}\sqrt{\pi_{n\gamma}}\sqrt{\pi_{n\eta}}a^*_{n\alpha|ij}a_{n\beta|i^{\prime}j}a^*_{n\beta|k^{\prime}l}a_{n\alpha|kl}a^*_{n\gamma|i^{\prime}j}a_{n\eta|ij}a^*_{n\eta|kl}a_{n\gamma|k^{\prime}l}.
\end{eqnarray}
as $i=p^{\prime}$ and $k=q^{\prime}$.

The entanglement in the $12|34$ bipartition of the post-measurement state when the $\Pi_n$ element of any arbitrary two-qubits measurement is performed on the (2,3) pair can be represented using Eq.\ref{eqa5}
\begin{eqnarray}
    \label{eqa6}
    \mathcal{C}_{n|12vs34}(|\Phi_n)\rangle\langle\Phi_n|)=\sqrt{\frac{4}{3}\left[1-\frac{1}{(\mbox{Tr}\Pi_n)^2}\sum_{\substack{\alpha,\beta,\\ \gamma,\eta=1}}^4\sum_{\substack{i,j,i^{\prime},k,\\l,k^{\prime}=0}}^1\sqrt{\pi_{n\alpha}}\sqrt{\pi_{n\beta}}\sqrt{\pi_{n\gamma}}\sqrt{\pi_{n\eta}}a^*_{n\alpha|ij}a_{n\beta|i^{\prime}j}a^*_{n\beta|k^{\prime}l}a_{n\alpha|kl}a^*_{n\gamma|i^{\prime}j}a_{n\eta|ij}a^*_{n\eta|kl}a_{n\gamma|k^{\prime}l}\right]}
\end{eqnarray}

\section{Negativity of the post-measurement state after sequential measurements}\label{ap3}

Negativity of a bipartite state $\rho_{AB}$ is  defined as \cite{vidal}
\begin{eqnarray}
    \label{eq23}
    \mathcal{N}(\rho_{AB})=\frac{\|\rho_{AB}^{T_B}\|-1}{2}
\end{eqnarray}
with $\|\rho_{AB}^{T_B}\|=\mbox{Tr}\sqrt{(\rho_{AB}^{T_B})^{\dagger}\rho_{AB}^{T_B}}$ where $T_B$ denote the partial transpose with respect to the subsystem $B$. We evaluate the entanglement of $\rho_{14|nm}$ (Eq.\ref{eq17}). 

Every orthonormal basis $|\phi_{n\alpha}\rangle$ and $|\psi_{m\delta}\rangle$ of different POVM elements can be written in the following computational basis form $|\phi_{n\alpha}\rangle=\sum_{i_1,j_1=0}^1a_{n\alpha|i_1j_1}|i_1j_1\rangle$ and $|\psi_{m\delta}\rangle=\sum_{l_1,k_1=0}^1b_{m\delta|l_1k_1}|l_1k_1\rangle$ where $\sum_{i_1,j_1=0}^1|a_{n\alpha|i_1j_1}|^2=1$, $a_{n\alpha|i_1j_1}\in\mathbb{C}$ and $\sum_{l_1,k_1=0}^1|b_{m\delta|l_1k_1}|^2=1$, $b_{m\delta|l_1k_1}\in\mathbb{C}$. So, using the above computational basis form, Eq.\ref{eq17} can be rewritten as:
\begin{eqnarray}
    \label{eqb1}
    \rho_{14|nm}=\frac{1}{4p_ns_{nm}}\sum_{\substack{\alpha,\beta,\zeta=1}}^4\sum_{\substack{i_{1,...,4},\\j_{1,...,4}=0}}^1\sum_{\substack{k_{1,...,2},\\l_{1,...,2}=0}}^1\mu_{m\zeta}\sqrt{\pi_{n\alpha}}\sqrt{\pi_{n\beta}}a_{n\alpha|i_1j_1}^*b_{m\zeta|k_1l_1}b^*_{m\zeta|k_2l_2}a_{n\beta|i_2j_2}a_{n\alpha|i_3j_3}a^*_{n\beta|i_4j_4}|i_3j_3\rangle\langle i_4j_4|
\end{eqnarray}
where $p_n=\sum_{\alpha=1}^4\frac{\pi_{n\alpha}}{4}$ and $s_{nm}=\frac{1}{p_n}\sum_{\alpha,\delta=1}^4\sum_{i,j=0}^1\pi_{n\alpha}\mu_{m\delta}|a_{n\alpha|ij}b^*_{m\delta|ij}|^2$.

The square root of any Hermitian, positive semidefinite matrix $U$ is given by \cite{horn}:
\begin{eqnarray}
    \label{eqb2}
 \sqrt{U}=\frac{U+\sqrt{\det(U)}\mathbb{I}}{\sqrt{\mbox{Tr}(U)+2\sqrt{\det (U)}}}
\end{eqnarray}

Let a matrix $U$ be defined as $U=(\rho_{14|nm}^{T_B})^{\dagger}\rho_{14|nm}^{T_B}$, where $T_B$ denotes the partial transpose with respect to qubit 4. Using the Eq.\ref{eqb1}, $U$ can be written as:

\begin{eqnarray}
    \label{eqb3}
    U&=&(\rho_{14|nm}^{T_B})^{\dagger}\rho_{14|nm}^{T_B}\nonumber\\
    &=&\frac{1}{16p_n^2s_{nm}^2}\sum_{\substack{\alpha,\beta,\zeta,\\ \gamma,\eta,\delta=1}}^4\sum_{\substack{i_{1,...,8},\\j_{1,...,8}=0}}^1\sum_{\substack{k_{1,...,4},\\l_{1,...,4}=0}}^1\mu_{m\zeta}\sqrt{\pi_{n\alpha}}\sqrt{\pi_{n\beta}}a_{n\alpha|i_1j_1}b^*_{m\zeta|k_1l_1}b_{m\zeta|k_2l_2}a^*_{n\beta|i_2j_2}a^*_{n\alpha|i_3j_3}a_{n\beta|i_4j_4}\nonumber\\
    &\times&\mu_{m\delta}\sqrt{\pi_{n\gamma}}\sqrt{\pi_{n\eta}}a^*_{n\gamma|i_5j_5}b_{m\delta|k_3l_3}b^*_{m\delta|k_4l_4}a_{n\eta|i_6j_6}a_{n\beta|i_3j_7}a^*_{n\eta|i_8j_4}|i_4j_3\rangle\langle i_8j_7| 
\end{eqnarray}
 We denote $X=\mbox{Tr}(U)$ and $Y=\det U$ which are given by:
\begin{eqnarray}
    \label{eqb4}
    X&=&\mbox{Tr}(U)\nonumber\\
    &=&\frac{1}{16p_n^2s_{nm}^2}\sum_{\substack{\alpha,\beta,\zeta,\\ \gamma,\eta,\delta=1}}^4\sum_{\substack{i_{1,...,8},\\j_{1,...,8}=0}}^1\sum_{\substack{k_{1,...,4},\\l_{1,...,4}=0}}^1\mu_{m\zeta}\sqrt{\pi_{n\alpha}}\sqrt{\pi_{n\beta}}a_{n\alpha|i_1j_1}b^*_{m\zeta|k_1l_1}b_{m\zeta|k_2l_2}a^*_{n\beta|i_2j_2}a^*_{n\alpha|i_3j_3}a_{n\beta|i_4j_4}\nonumber\\
    &\times&\mu_{m\delta}\sqrt{\pi_{n\gamma}}\sqrt{\pi_{n\eta}}a^*_{n\gamma|i_5j_5}b_{m\delta|k_3l_3}b^*_{m\delta|k_4l_4}a_{n\eta|i_6j_6}a_{n\beta|i_3j_3}a^*_{n\eta|i_4j_4}
\end{eqnarray}
and 
\begin{eqnarray}
    \label{eqb5}
   Y= \det U=\sum_{e,f,g,h=1}^4 \epsilon_{efgh}\tau_{1e}\tau_{2f}\tau_{3g}\tau_{4h}
\end{eqnarray}
where $\epsilon_{efgh}$ is Levi-Civita symbol. The row elements of the matrix $(\rho_{14|nm}^{T_B})^{\dagger}\rho_{14|nm}^{T_B}$ can be denoted as $\tau_{1e}, \tau_{2f}, \tau_{3g}, \tau_{4h}$ for $e,f,g,h \in {1,2,3,4}$, where the first, second, third and fourth rows respectively. Using these elements, by setting $i_4=0$ and $j_3=0$ in Eq.\ref{eqb3}, we  obtain
\begin{eqnarray}
    \label{eqb6}
    \tau_{1e}&=&\frac{1}{16p_n^2s_{nm}^2}\sum_{\substack{\alpha,\beta,\zeta,\\ \gamma,\eta,\delta=1}}^4\sum_{\substack{i_{1,...,8},\\j_{1,...,8}=0}}^1\sum_{\substack{k_{1,...,4},\\l_{1,...,4}=0}}^1\mu_{m\zeta}\sqrt{\pi_{n\alpha}}\sqrt{\pi_{n\beta}}a_{n\alpha|i_1j_1}b^*_{m\zeta|k_1l_1}b_{m\zeta|k_2l_2}a^*_{n\beta|i_2j_2}a^*_{n\alpha|i_30}a_{n\beta|0j_4}\nonumber\\
    &\times&\mu_{m\delta}\sqrt{\pi_{n\gamma}}\sqrt{\pi_{n\eta}}a^*_{n\gamma|i_5j_5}b_{m\delta|k_3l_3}b^*_{m\delta|k_4l_4}a_{n\eta|i_6j_6}a_{n\beta|i_3j_7}a^*_{n\eta||i_8j_4}
\end{eqnarray}
Similarly, by setting $i_4=0$  and $j_3=1$ 
in Eq.\ref{eqb3}, we get
\begin{eqnarray}
    \label{eqb7}
    \tau_{2f}&=&\frac{1}{16p_n^2s_{nm}^2}\sum_{\substack{\alpha,\beta,\zeta,\\ \gamma,\eta,\delta=1}}^4\sum_{\substack{i_{1,...,8},\\j_{1,...,8}=0}}^1\sum_{\substack{k_{1,...,4},\\l_{1,...,4}=0}}^1\mu_{m\zeta}\sqrt{\pi_{n\alpha}}\sqrt{\pi_{n\beta}}a_{n\alpha|i_1j_1}b^*_{m\zeta|k_1l_1}b_{m\zeta|k_2l_2}a^*_{n\beta|i_2j_2}a^*_{n\alpha|i_31}a_{n\beta|0j_4}\nonumber\\
    &\times&\mu_{m\delta}\sqrt{\pi_{n\gamma}}\sqrt{\pi_{n\eta}}a^*_{n\gamma|i_5j_5}b_{m\delta|k_3l_3}b^*_{m\delta|k_4l_4}a_{n\eta|i_6j_6}a_{n\beta|i_3j_7}a^*_{n\eta||i_8j_4},
\end{eqnarray}
  and by putting $i_4=1$ and $j_3=0$ in Eq.\ref{eqb3} we get
 
\begin{eqnarray}
    \label{eqb8}
    \tau_{3g}&=&\frac{1}{16p_n^2s_{nm}^2}\sum_{\substack{\alpha,\beta,\zeta,\\ \gamma,\eta,\delta=1}}^4\sum_{\substack{i_{1,...,8},\\j_{1,...,8}=0}}^1\sum_{\substack{k_{1,...,4},\\l_{1,...,4}=0}}^1\mu_{m\zeta}\sqrt{\pi_{n\alpha}}\sqrt{\pi_{n\beta}}a_{n\alpha|i_1j_1}b^*_{m\zeta|k_1l_1}b_{m\zeta|k_2l_2}a^*_{n\beta|i_2j_2}a^*_{n\alpha|i_30}a_{n\beta|1j_4}\nonumber\\
    &\times&\mu_{m\delta}\sqrt{\pi_{n\gamma}}\sqrt{\pi_{n\eta}}a^*_{n\gamma|i_5j_5}b_{m\delta|k_3l_3}b^*_{m\delta|k_4l_4}a_{n\eta|i_6j_6}a_{n\beta|i_3j_7}a^*_{n\eta||i_8j_4},
\end{eqnarray}
lastly, putting $i_4=1$ and $j_3=1$ in Eq.\ref{eqb3} we get
\begin{eqnarray}
    \label{eqb9}
    \tau_{4h}&=&\frac{1}{16p_n^2s_{nm}^2}\sum_{\substack{\alpha,\beta,\zeta,\\ \gamma,\eta,\delta=1}}^4\sum_{\substack{i_{1,...,8},\\j_{1,...,8}=0}}^1\sum_{\substack{k_{1,...,4},\\l_{1,...,4}=0}}^1\mu_{m\zeta}\sqrt{\pi_{n\alpha}}\sqrt{\pi_{n\beta}}a_{n\alpha|i_1j_1}b^*_{m\zeta|k_1l_1}b_{m\zeta|k_2l_2}a^*_{n\beta|i_2j_2}a^*_{n\alpha|i_31}a_{n\beta|1j_4}\nonumber\\
    &\times&\mu_{m\delta}\sqrt{\pi_{n\gamma}}\sqrt{\pi_{n\eta}}a^*_{n\gamma|i_5j_5}b_{m\delta|k_3l_3}b^*_{m\delta|k_4l_4}a_{n\eta|i_6j_6}a_{n\beta|i_3j_7}a^*_{n\eta||i_8j_4}.
\end{eqnarray}

So, the negativity of the final post-measurement state $\rho_{14|nm}$ (Eq.\ref{eq17}) is
\begin{eqnarray}
    \label{eq28}
    \mathcal{N}(\rho_{14|nm})&=&\frac{\mbox{Tr}\sqrt{U}-1}{2}\nonumber\\
    &=&\frac{X+4\sqrt{Y}}{2\sqrt{X+2\sqrt{Y}}}-\frac{1}{2}
\end{eqnarray}
where $X$ and $Y$ are given above in Eq.\ref{eqb4} and Eq.\ref{eqb5} respectively. We get a relation between the negativity of the post-measurement state and the measurement parameters of two successive measurements.

\twocolumngrid

\end{document}